\documentclass[aps,prl,twocolumn,nopacs,superscriptaddress]{revtex4-2}
\usepackage{epsfig,mathrsfs,xcolor,latexsym,subfigure,marginnote,graphicx,verbatim,relsize,mathrsfs,color,array,amsmath,amsfonts,amssymb,graphicx}
\usepackage{braket}
\usepackage[english]{babel}
\usepackage[utf8]{inputenc}
\usepackage[colorinlistoftodos, color=green!40, prependcaption]{todonotes}
\usepackage{amsthm}
\usepackage{mathtools}
\usepackage{physics}
\usepackage{xcolor}
\usepackage{graphicx}
\usepackage[normalem]{ulem} 
\usepackage{adjustbox}
\usepackage{placeins}
\usepackage[T1]{fontenc}
\usepackage{lipsum}
\usepackage{csquotes}
\usepackage{float}
\usepackage{booktabs}
\usepackage{amsmath}
\usepackage{float}
\usepackage{array}
\usepackage[breaklinks=true,linkcolor=blue, pdftex, pdftitle={Article}, pdfauthor={Author}, colorlinks=true, citecolor=blue, urlcolor=blue, filecolor=blue]{hyperref}
\usepackage{cleveref}
\usepackage{chemformula}

\newcommand{\tl}[1]{{\color{red} #1}}

\newcommand{\mx}{\mathrm{x}}
\newcommand{\mc}{\mathrm{c}}

\newcommand{\mH}{\mathrm{H}}
\newcommand{\vecr}{\textbf{r}}

\begin{document}
\title{Artificial Symmetry Breaking by Self-Interaction Error
}

\author{Lin Hou}
\author{Cody Woods}
\author{Yanyong Wang}
\author{Jorge Vega Bazantes}
\author{Ruiqi Zhang}
\affiliation{
Department of Physics and Engineering Physics, Tulane University, New Orleans, Louisiana 70118, USA}
\author{Shimin Zhang}
\author{Erik Alfredo Perez Caro}
\author{Yuan Ping}
\affiliation{Department of Materials Science and Engineering, University of Wisconsin-Madison, 53706, USA}
\author{Timo Lebeda}
\email[]{tlebeda@tulane.edu}
\author{Jianwei Sun}
\email[]{jsun@tulane.edu}
\affiliation{
Department of Physics and Engineering Physics, Tulane University, New Orleans, Louisiana 70118, USA}
\date{\today} 

\begin{abstract}

Symmetry is a cornerstone of quantum mechanics and materials theory, underpinning the classification of electronic states and the emergence of complex phenomena such as magnetism and superconductivity. While symmetry breaking in density functional theory can reveal strong electron correlation, it may also arise spuriously from self-interaction error (SIE), an intrinsic flaw in many approximate exchange-correlation functionals. In this work, we present clear evidence that SIE alone can induce artificial symmetry breaking, even in the absence of strong correlation. Using a family of one-electron, multi-nuclear-center systems \( \mathrm{H}^+_{n \times \frac{+2}{n}}(R) \), we show that typical semilocal density functionals exhibit symmetry-breaking localization as system size increases, deviating from the exact, symmetry-preserving Hartree-Fock solution. We further demonstrate that this localization error contrasts with the well-known delocalization error of semilocal density functionals and design a semilocal density functional that avoids the artifact. Finally, we illustrate the real-world relevance of this effect in the \ch{Ti_{Zn}v_O} defect in ZnO, where a semilocal density functional breaks the $C_{3v}$ symmetry while a hybrid density functional preserves it. These findings highlight the need for improved functional design to prevent spurious symmetry breaking in both model and real materials.

\end{abstract}

\maketitle
Symmetry plays a foundational role in quantum mechanics, condensed matter physics, chemistry, and materials science. It governs conservation laws, constrains the form of the Hamiltonian, and shapes the structure of the Hilbert space. In many-body systems, symmetry not only guides theoretical formulations but also enables the classification of phases and the prediction of electronic, magnetic, and optical properties. Importantly, subtle symmetry breaking is often associated with the emergence of novel physical phenomena such as superconductivity~\cite{bardeen1957theory,schrieffer2018theory,tinkham2004introduction,ginzburg2009theory}, ferromagnetism~\cite{sadler2006spontaneous}, and topological phases~\cite{nakahara2018geometry}. As such, symmetry serves both as a constraint and a powerful diagnostic tool for exploring the underlying physics of interacting electrons.

In many cases, strong electron correlation arises from degeneracies or near-degeneracies dictated by symmetry. A paradigmatic example is the hydrogen molecule at stretched bond lengths: although the exact ground state remains a spin singlet, the system becomes increasingly statically correlated due to the near-degeneracy of two configurations with electrons localized on opposite nuclei. This type of strong correlation poses a major challenge for Density Functional Theory (DFT) \cite{hohenberg1964inhomogeneous}, whose widely used approximations typically rely on single-determinant references and struggle to capture the essential multi-reference character. In such situations, symmetry breaking in approximate DFT can provide a practical workaround, allowing the system to escape the limitations of single-reference descriptions and partially recover the missing correlation energy~\cite{perdew2021interpretations,zhang2020symmetry,gunnarsson1976exchange}.

The theoretical foundations and practical benefits of symmetry breaking within DFT have been explored extensively. Many work have shown that symmetry-broken solutions obtained with standard approximate functionals can yield physically meaningful results in strongly correlated systems, including complex solids~\cite{zhang2020symmetry,zhang2020competing,zhang2024emergence,trimarchi2018polymorphous,varignon2019origin,zunger2022bridging}. Perdew and collaborators have provided an alternative interpretation, viewing symmetry-broken spin-density functionals as approximations to a more accurate energy functional that includes the on-top pair density~\cite{perdew1997top}. These perspectives highlight symmetry breaking not merely as an artifact but as a useful tool for revealing underlying physical effects, for example in antiferromagnetism~\cite{perdew2021interpretations} and in cases like $H_2$ at bond dissociation where symmetry-preserving DFT fails to capture correct energetics or density distributions~\cite{gunnarsson1976exchange}.

However, symmetry breaking in DFT is not without risk. Another long-standing limitation of approximate functionals is the \emph{self-interaction error} (SIE)---the spurious interaction of an electron with itself due to incomplete cancellation of the Hartree and exchange contributions~\cite{perdew1981self}. SIE leads to a wide range of qualitative and quantitative failures, including incorrect dissociation limits, underestimated energy barriers, underestimated ionization potentials, underestimated band gaps in insulators, and overly delocalized charge distributions~\cite{perdew1981self,ruzsinszky2006spurious,mori2006many,kronik2020piecewise}. In the presence of SIE, symmetry breaking may yield misleading or unphysical results. Indeed, Zunger and his colleagues has cautioned that symmetry breaking should only be applied when using functionals with minimal SIE
to ensure that the resulting broken-symmetry states reflect real physical behavior rather than artifacts of the functional form~\cite{zhang2020symmetry, trimarchi2018polymorphous}. 
For instance, in point defects within wide band gap materials—promising candidates for qubits—preserving point group symmetry is often essential for maintaining correct electronic and spin properties~\cite{karapatzakis2024microwave, zhang2025transition}, making any symmetry breaking induced by SIE particularly undesirable as we will show later.   Yet, to date, there has been no definitive demonstration that SIE alone can spuriously drive symmetry breaking. 

In this work, we present clear and direct evidence that \emph{self-interaction error can spuriously drive symmetry breaking}. To this end, we design a \emph{one-electron model system}
in which, by definition, there is no strong correlation and for which Hartree-Fock (HF) theory provides an exact reference. In this one-electron model system HF therefore
preserves the full symmetry of the Hamiltonian and any symmetry breaking in a DFT solution must be attributed to the approximation itself---specifically to SIE. 

Beyond the one-electron regime, approximate functionals also suffer from \emph{many-electron self-interaction errors} \cite{mori2006many}, which manifest as \emph{delocalization errors}. These errors are characterized by convex deviations from the piecewise linear behavior of the total energy $E(N)$ with respect to fractional electron number~\cite{mori2008localization,bryenton2023delocalization}, leading to systematic errors in band gaps, reaction barriers, polarizabilities, the dissociation of molecular ions, charge-transfer states, and ionization potentials and electron affinities \cite{cohen2008insights}. In contrast, Hartree-Fock tends to exhibit a \emph{localization error}, associated with its concave energy profile. Recent theoretical analysis by Chen and Yang~\cite{li2017piecewise} challenges the prevailing assumption that semilocal density functional approximations (DFAs) are always convex. They demonstrated that such functionals can exhibit narrow concave regions in $E(N)$, leading to unexpected and potentially unphysical localization of electron density across subsystems.

Building on this insight, we validate Chen and Yang’s conjecture by showing that this narrow concavity in typical semilocal functionals leads to \emph{artificial symmetry breaking} in the one-electron model system. Specifically, we generalize H$_2^+$---a benchmark for probing SIE---into a family of one-electron, multi-nuclear-center systems denoted $\mathrm{H}^+_{n \times \frac{+2}{n}}(R)$. For these systems, HF accurately delocalizes the electron across all centers, preserving the global symmetry. However, semilocal DFAs can break this symmetry and spuriously localize the electron on a subset of centers. These results underscore the critical importance of reducing self-interaction and delocalization errors in functional design. As a proof of concept, we construct a semilocal DFA that avoids artificial symmetry breaking and maintains accurate electron delocalization in the test systems.



\begin{figure}[htb]
    \centering
    \includegraphics[width=\linewidth]{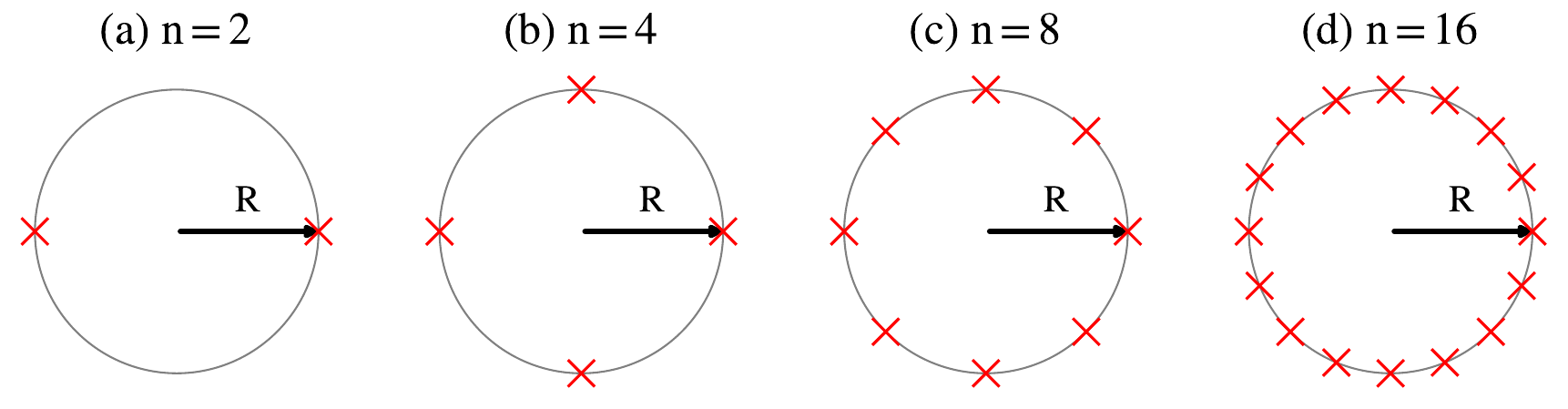}
    \caption{Definition of the multi-nuclear-center one-electron model \( \mathrm{H}^+_{n \times \frac{+2}{n}}(R) \). The panels (a)-(d) illustrate configurations with \( n = 2, 4, 8, \) and \( 16 \) nuclear centers. The nuclei, denoted by crosses, are evenly distributed on a circle of radius \( R \). Each nucleus carries a fractional charge of \( +\frac{2}{n}e \). Thus, the total nuclear charge sums to \( +2e \), and the system carries a net charge of \( +e \). The shortest internuclear distance is $d=2R\sin(\pi/n)$.
    }
    \label{fig:model}
\end{figure}

In Fig.~\ref{fig:model}, we define the \( \mathrm{H}^+_{n \times \frac{+2}{n}}(R) \) model that provides a rich platform for exploring the influence of both the number of nuclear centers \( n \) and the radius \( R \) on the SIE. Despite the model's simplicity, it presents a significant challenge for DFT: 
Using the local density approximation (LDA), the Perdew-Burke-Ernzerhof functional (PBE) \cite{perdew1996generalized}, and the strongly constrained and appropriately normed (SCAN) functional \cite{sun2015strongly} as examples, we demonstrate that typical semilocal DFAs exhibit an artificial symmetry breaking in the electron density as the number of non-integer charged centers and the radius increase.
Except for those for Fig.~\ref{fig:defect}, all DFT calculations throughout this work were performed in \textsc{Turbomole} \cite{Turbomole2020,Weigend1998_RIJ,Neese2003_RIK} using the d-aug-cc-pVQZ basis set \cite{Dunning1989}.

\begin{figure}[htb]
    \centering
    \includegraphics[width=0.99\linewidth]{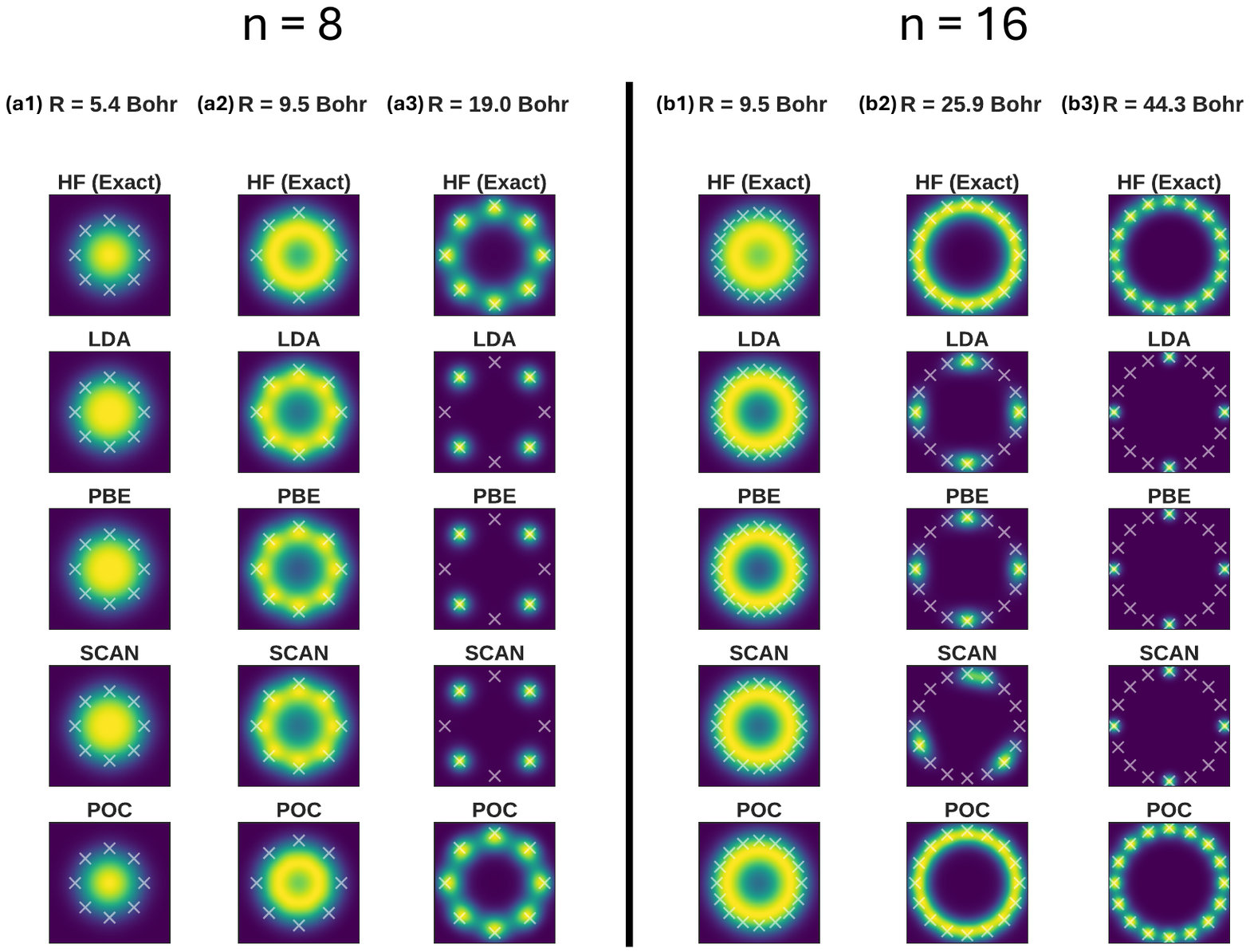}
    \caption{Electron density distribution for selected radii \( R \) for the \( \mathrm{H}^+_{n \times \frac{+2}{n}}(R) \) model with \( n=8 \) (left) and \( n=16 \) (right). Results from LDA, PBE, SCAN, and POC compared with the HF reference. The nuclei are located at the positions marked by white \( \times \), with charges of \( +0.25e \) for n=8 and \( +0.125e \) for n=16. Brighter color means higher electron density and the color scale is consistent within each column.
}
\label{fig:Edens}
\end{figure}

Figure~\ref{fig:Edens} shows that HF and the proof-of-concept (POC) DFA introduced below preserve the global symmetry of the electron density across all nuclear centers and regardless of the radius \( R \). In contrast, LDA, PBE, and SCAN break the global \( \text{C}_8 \) and \( \text{C}_{16} \) symmetries of the electron density for n=8 and n=16, respectively, as \( R \) increases. These typical semilocal DFAs exhibit a preference for electron localization around only four centers at large radii for \( n=8 \) and \( n=16 \). Interestingly, for n=16 at intermediate radius $R = 24.86$ Bohr, the SCAN functional localizes the electron density predominantly in three regions -- one of which is spread across two nuclear centers. Thus, for \( n=8 \) and \( n=16 \), typical semilocal DFAs suffer from a \emph{localization error}, in contrast to their usual delocalization error.

For smaller $R$ both the semilocal DFAs and HF exhibit ``donut'' shaped electron densities that preserve the global symmetry. Notably, all the electron density ``donut'' do not lie precisely on the circle where the nuclei reside. The HF and POC densities form a much smaller ``donut'' compared to the typical semilocal DFAs, reflecting the delocalization error of the latter. This delocalization error is more pronounced for n=8 than for n=16.

It is interesting to note that for small radii the typical semilocal DFAs show a delocalization error, see Fig.~\ref{fig:Edens}(a1) and (a2), whereas for large radii they show a localization error, see Fig.~\ref{fig:Edens}(a3). Remarkably, POC not only removes the localization error in Fig.~\ref{fig:Edens}(a3), but also substantially reduces the delocalization error in Fig.~\ref{fig:Edens}(a1 and a2), even leading to a slight overlocalization. For $n=16$, however, POC only gets rid of the localization at large radii, but not entirely of the delocalization in Fig.~\ref{fig:Edens}(b1).

In the Supplemental Material \cite{supplemental}, we provide results for a continuous variation of \( R \). There we observe that artificial symmetry breaking occurs almost gradually with increasing \( R \) and only for \( n > 4 \). For \( n=8 \), the electron density on the nuclear sites at 12, 3, 6, and 9 o'clock gradually decreases, and eventually vanishes, as \( R \) grows. We observe a similar trend for \( n=16 \), although for SCAN the transition from three-center to four-center occupation is abrupt, indicating a degeneracy of the three-center and four-center states at $R=32.5$ Bohr.

The mechanism that drives this artificial symmetry breaking can be rationalized from the analysis of Li and Yang \cite{li2017piecewise} for the curvature of $E(N)$, i.e., of the total energy with respect to the particle number. For the exact functional, this curvature vanishes, leading to the well-known piecewise linearity condition \cite{perdew1982density}. For semilocal DFAs, the curvature is usually positive, resulting in the famous delocalization error \cite{bryenton2023delocalization}. However, for small fractional charge, that is $N=N_0+\delta$ with $N_0$ an integer and $0\leq \delta\ll1$, the curvature can become negative and eventually tend to $-\infty$ as $\delta\to0^+$ \cite{li2017piecewise}.
In such regions of concave $E(N)$, semilocal DFAs energetically favor localized electron density distributions over delocalized ones, just like HF does for many-electron systems. The artificial symmetry breaking associated with electron density localization thus arises naturally from the inability of semilocal DFAs to achieve the piecewise linearity of $E(N)$.

An instructive limit to understand the origin of the artificial symmetry breaking and the associated localization error is the dissociation limit, in which SIE is usually most pronounced. In the following, we consider the more general case of adding or removing a fractional amount of charge to an atom.
Let $\rho_0$ denote the electron density of the isolated atom and assume that the electron density after adding or removing charge can be approximated by $\rho_\delta= (1+\delta) \rho_0$ for some $-1<\delta<1$. Then the reduced density gradient, ${s=\vert\nabla \rho\vert / [2(3\pi^2)^{1/3} \rho^{4/3} ]}$, becomes $s_\delta = (1+\delta)^{1/3}s_0$, with $s_0$ the reduced density gradient of the isolated atom. The exact condition for freedom from SIE is \cite{perdew1981self}
\begin{equation} \label{eq:SIC condition}
    E_\mx[\rho] + E_\mH[\rho] = 0~, \quad E_\mc[\rho] = 0
\end{equation}
for all one-electron densities $\rho$, where \( E_\mH \) is the Hartree energy, \( E_\mx \) is the exchange energy, and \( E_\mc \) is the correlation energy.
In the following, we neglect the condition on correlation, as the latter can be easily satisfied in semilocal DFAs that use the kinetic energy density \cite{becke1988correlation}. For the Hartree energy of $\rho_\delta$, we obtain
\begin{align} \label{eq:Hartree-scaling}
    E_\mH[\rho_\delta] &= -\frac12 \iint \frac{\rho_\delta(\vecr) \rho_\delta(\vecr')}{\vert \vecr-\vecr'\vert} d^3r' d^3r \nonumber \\
    &= (1+\delta)^{2} E_\mH[\rho_0]~.
\end{align}
The exchange energy of a GGA is usually defined in terms of its enhancement factor $F_\mx$ as
\begin{equation}
    E_\mx[\rho] = A_\mx \int \rho^{4/3}(\vecr) F_\mx(s) d^3r~,
\end{equation}
where $A_\mx = - (3/4) \left(3/\pi \right)^{1/3}$. Using $\rho_\delta= (1+\delta)\rho_0$ yields $E_\mx[\rho_\delta] = (1+\delta)^{4/3} A_\mx \int \rho_0^{4/3}(\vecr) F_\mx(s_\delta) d^3r$. Matching the scaling of the Hartree energy, Eq.~\eqref{eq:Hartree-scaling}, thus requires
\begin{equation} \label{eq:SIC-scaling Fx}
    F_\mx(s_\delta) = (1+\delta)^{2/3} F_\mx(s_0)~.
\end{equation}
Recalling $s_\delta = (1+\delta)^{-1/3}s_0$, condition \eqref{eq:SIC-scaling Fx} is satisfied if
\begin{equation} \label{eq:SIC-condition for Fx}
    F_\mx\propto s^{-2}~.
\end{equation}
In the case of \( \mathrm{H}^+_{n \times \frac{+2}{n}}(R\to\infty) \), $\rho_0$ is the electron density of an isolated hydrogen atom of nuclear charge $Z=2/n$, denoted by $\rho_{\mH(Z)}$. The electron density around each nuclear center, $\rho_\delta = \rho_{\mH(Z)}/n =: \rho_{\mH/n(Z)}$, satisfies the condition ${\rho_\delta= (1+\delta) \rho_0}$ for $\delta = 1/n - 1$, i.e., the derivation is exact. By virtue of the uniform density scaling \cite{levy1985hellmann}, the dependence on fractional nuclear charge can also be dropped, and condition \eqref{eq:SIC-scaling Fx} becomes $F_\mx(s_{\mH/n(Z)}) = n^{-2/3} F_\mx(s_\mH)$ with $s_\mH$ the reduced density gradient of a hydrogen atom, as we detail in the Supplemental Material \cite{supplemental}.

Equations~\eqref{eq:SIC-scaling Fx} and \eqref{eq:SIC-condition for Fx} provide a semilocal realization of the exact constraint \eqref{eq:SIC condition} in the dissociation limit of \( \mathrm{H}^+_{n \times \frac{+2}{n}}(R\to\infty) \) and thus allow to remove SIE in this limit. We rationalize in the Supplemental Material \cite{supplemental,singh2019adiabatic,SORFKL} why this condition also removes the region of negative curvature.

In practice, Eq.~\eqref{eq:SIC-condition for Fx} can not be satisfied for all $s$. The divergence for $s\to0$ is not only undesirable from a physical point of view, but also leads to numerical instability. As Eq.~\eqref{eq:SIC-condition for Fx} further contradicts the condition $F_\mx\sim s^{-1/2}$ as $s\to\infty$ that guarantees a more correct limit under nonuniform coordinate scaling to the true two-dimensional limit \cite{levy1993tight,sun2015strongly}, we have to seek a compromise between satisfying Eq.~\eqref{eq:SIC-condition for Fx} in the energetically important region and fulfilling other exact constraints, as well as numerical stability. As a proof-of-concept (POC), we propose the simple form
\begin{equation}
    F_\mx^\mathrm{POC}(s) = \frac{h_\mx^0}{1+(s/a)^2}~.
\end{equation}
We satisfy the tight bound for two-electron systems \cite{strongly_tightened_bound} by $h_\mx^0=1.174$. As a compromise between satisfying Eq.~\eqref{eq:SIC-condition for Fx} in the energetically important region and numerical stability, we choose $a=2$. We also investigated fixing $a$ by the exchange energy of the hydrogen atom, which yields $a=12.37$. However, the resulting functional does not fulfill condition \eqref{eq:SIC-condition for Fx} in the energetically important region, but only in regions far from the nucleus.

We emphasize that the derivation of condition \eqref{eq:SIC-condition for Fx} holds for many-electron systems as well. However, ${E_\mx+E_\mH=0}$ is an exact constraint only for one-electron densities and usually not desirable for many-electron densities. Thus, condition \eqref{eq:SIC-condition for Fx}, and accordingly also using POC, is only justified in one-electron regions.

\begin{figure}[htb]
    \centering
    \includegraphics[width=\linewidth]{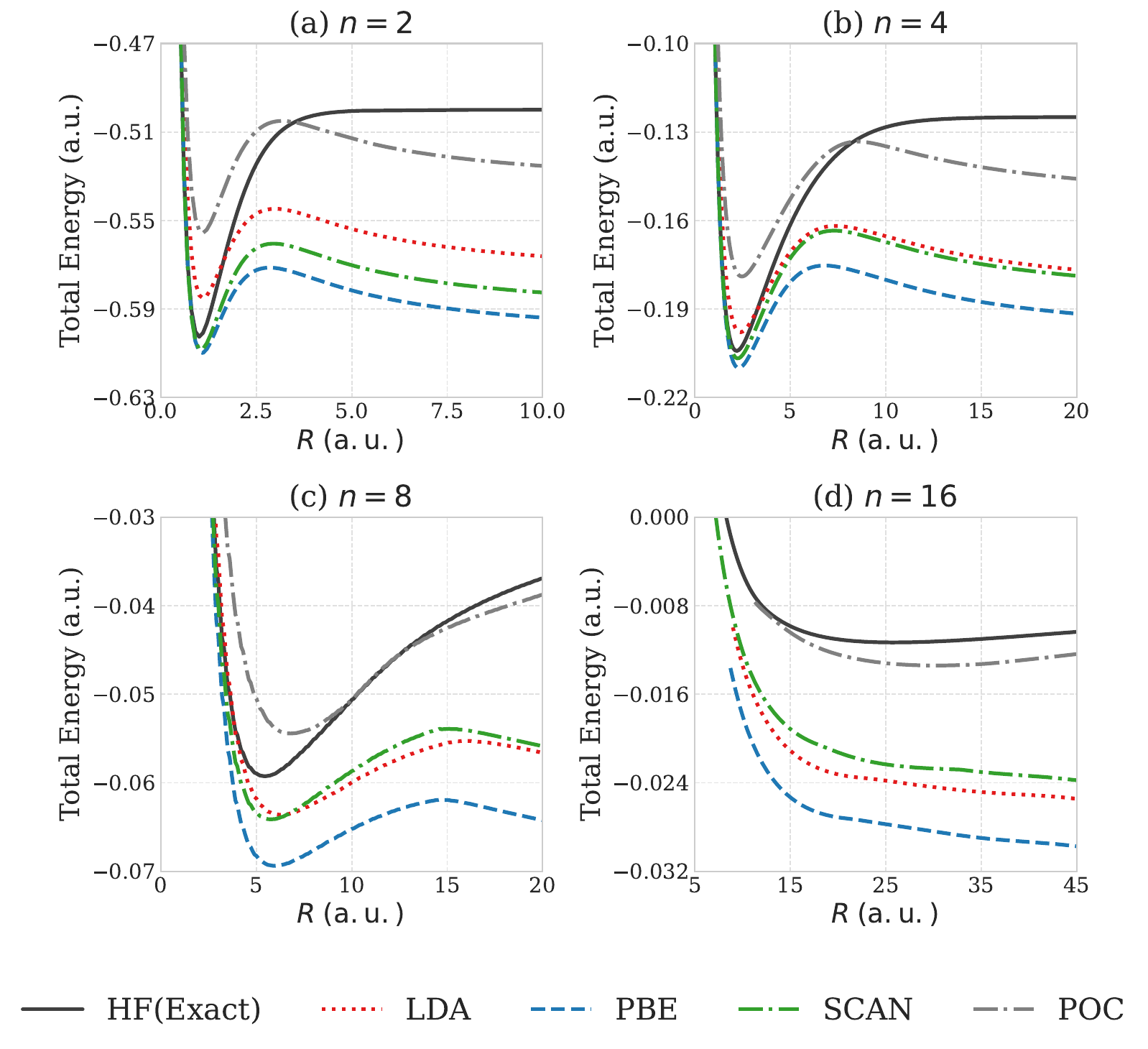}
    \caption{Total energy of the multicenter one-electron system $\mathrm{H}^+_{n \times \frac{+2}{n}}(R)$ for $n = 2, 4, 8, 16$, respectively. 
    }
    \label{fig:E_bind_err}
\end{figure}

To further study the process of this artificial symmetry-breaking phenomenon, we compare the total energies obtained with LDA, PBE, SCAN, POC, and HF for \( \mathrm{H}^+_{n \times \frac{+2}{n}}(R) \) across different values of \( n \) and \( R \) in Fig.~\ref{fig:E_bind_err}. Since the HF reference is free of SIE, the deviations in DFAs can be attributed solely to SIE.

Near equilibrium, the typical smilocal DFAs provide reasonable approximations for \( n=2,~4, \) and \( 8 \). However, none of them predicts a stable equilibrium state for \( n=16 \). This marked failure underscores the limitation of the currently available semilocal functionals in describing extended multicenter systems.
As POC does not satisfy the hydrogen atom norm, limited accuracy near equilibrium is anticipated. Figure \ref{fig:E_bind_err} confirms this for \( n=2,~4, \) and \( 8 \). For n=16, however, POC clearly outperforms the other semilocal DFAs across the entire range.

As the radius \( R \) increases far beyond the HF equilibrium region, the discrepancy in total energy between the typical semilocal DFAs and the exact HF reference grows for all \( n \), highlighting the intrinsic deficiency of the former in describing molecular dissociation due to SIE.
In contrast, POC performs much better for intermediate and large radii across all $n$. Since SIE becomes increasingly important with increasing R, this indicates that POC achieves reduced SIE.

These results highlight that while typical semilocal DFAs capture energy trends around equilibrium, SIE significantly compromises their accuracy in stretched bond settings, particularly in highly symmetric, multicenter configurations. This urges the need for advanced functionals that better preserve electron density symmetry and minimize SIE in extended systems. POC provides a first step in this direction by reducing SIE in extended systems, although at the cost of accuracy for equilibrium properties.

While SCAN utilizes the hydrogen exchange energy to minimize SIE near equilibrium, POC is based on condition \eqref{eq:SIC-condition for Fx} to minimize SIE in the dissociation limit. It is not surprising that achieving both in a single semilocal DFA is challenging, as this requires modeling the nonlocal double integral in $E_\mH$.
Thus, despite recent advancements in semilocal functional design \cite{aschebrock2019ultranonlocality,furness2020accurate,mejia2020meta,lebeda2023right,lebeda2024balancing,lebeda2025meta,giri2025isostructural,SKALAluise2025accurate}, constructing a general purpose DFA using condition \eqref{eq:SIC-condition for Fx} for one-electron regions remains a major challenge. Not only does condition \eqref{eq:SIC-condition for Fx} not remove SIE near equilibrium, it is also in conflict with other exact constraints and entails numerical difficulties. One way out of this dilemma could be to include the Laplacian of the density that, in contrast to the kinetic energy density, can provide additional information in iso-orbital systems \cite{ramasamy2024semilocal}.

\begin{figure}[htb]
    \centering
    \includegraphics[width=\linewidth]{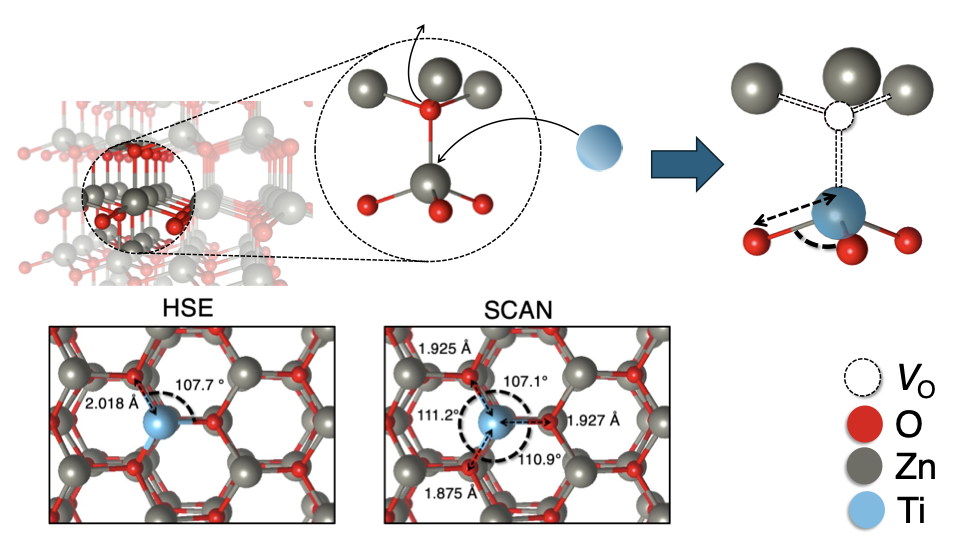}
    \caption{Top: Illustration of the \ch{Ti_{Zn}v_O} defect in \ch{ZnO}. Bottom: Geometry of \ch{Ti_{Zn}v_O} using HSE and SCAN. All calculations performed in \textsc{vasp} \cite{kresse1993ab,kresse1996efficient,kresse1999ultrasoft} and see computational details in Ref.~\cite{zhang2025transition}. }
    \label{fig:defect}
\end{figure}

To demonstrate the relevance of this SIE-induced symmetry breaking in real materials, we examine the relaxed geometries of the defect complex \ch{Ti_{Zn}v_O}---a Ti-on-Zn antisite with an adjacent oxygen vacancy in ZnO---calculated using the Heyd-Scuseria-Ernzerhof (HSE) hybrid functional~\cite{heyd2003hybrid} with a Fock exchange fraction of 0.375 and a screening parameter \( \omega = 0.2~\text{\AA}^{-1} \), which reproduces the experimental lattice constants and band gap~\cite{zhang2025transition}. As shown in Fig.~\ref{fig:defect}, HSE predicts the \( C_{3v} \) symmetry of the defect, which is often essential for maintaining the triplet ground state relevant for qubit functionality~\cite{zhang2025transition}. In contrast, SCAN lowers the symmetry to \( C_{1h} \). Since HSE significantly reduces SIE through its inclusion of screened exact exchange~\cite{zhang2019subtlety}, this discrepancy suggests that the symmetry breaking observed with SCAN is artificial and a direct consequence of its higher SIE, whereas the symmetry preserved by HSE is more likely to reflect the true physical ground-state structure.

In summary, we demonstrated that self-interaction error (SIE) alone can drive artificial symmetry breaking in density functional approximations. Using a one-electron model system \( \mathrm{H}^+_{n \times \frac{+2}{n}}(R) \), we showed that typical semilocal DFAs incorrectly localize the electron as system size increases, in contrast to the exact, symmetry-preserving Hartree-Fock solution—revealing a localization error distinct from the usual delocalization error. We proposed a semilocal DFA that avoids this artifact and, extending to real materials, showed that the SCAN functional breaks the symmetry of the \ch{Ti_{Zn}v_O} defect in ZnO, while HSE06 preserves it—supporting the conclusion that SIE can lead to spurious symmetry breaking in both model and practical systems.

\section*{Acknowledgements} \label{sec:acknowledgements}
This work was supported by National Science Foundation (NSF) under Grant No. DMR-2042618. JVB and RZ acknowledge the support of the U.S.
Department of Energy (DOE), Office of Science (OS), Basic
Energy Sciences (BES), Grant No. DE-SC0014208. by LH and CW acknowledge Henry Fitzhugh and Jamin Kidd for assisting with \textsc{Turbomole} plots.

\section*{Author contributions} 
LH and CW performed the calculations corresponding to figures 1 to 3. YW performed the calculations corresponding to figure 4 with help from RZ, SZ, EC, and YP. JVB prepared figure 4. LH, CW, and TL analyzed the results. TL conceived and implemented the proof-of-concept density functional. JS conceptualized the work and provided computational infrastructure. LH prepared figures 1 through 3 and wrote the first draft of the manuscript with help from TL. TL and JS worked out the final version. All authors discussed the final version.

\appendix*
\bibliography{ASBSIE}

\end{document}